\documentclass[twocolumn,showpacs,prl,floatfix,letter]{revtex4-1}
\ifx\pdfoutput\undefined
\usepackage{graphicx}
\else
\usepackage{epstopdf}
\usepackage{epsfig}
\usepackage{amssymb,amsmath,stmaryrd,tabularx}
\usepackage{wrapfig}
\usepackage{bm}
\fi

\usepackage[center]{subfigure}

\begin{document}
\title{Distribution of maximum velocities in avalanches near the depinning transition}
\author{Michael LeBlanc$^{1}$, Luiza Angheluta$^{1,2}$, Karin Dahmen$^{1}$ and Nigel Goldenfeld$^{1}$}
\affiliation{
$^1$Department of Physics, University of Illinois at
Urbana-Champaign, Loomis Laboratory of Physics, 1110 West Green
Street, Urbana, Illinois, 61801-3080\\
$^2$Physics of Geological Processes, Department of Physics, University of Oslo, Norway
}



\begin{abstract}
We report exact predictions for universal scaling exponents and scaling
functions associated with the distribution of the maximum collective
avalanche propagation velocities $v_m$ in the mean field theory of the
interface depinning transition. We derive the extreme value
distribution $P(v_m|T)$ for the maximum velocities in avalanches of
fixed duration $T$, and verify the results by numerical simulation near
the critical point. We find that the tail of the distribution of
maximum velocity for an arbitrary avalanche duration, $v_m$, scales as
$P(v_m)\sim v_m^{-2}$ for large $v_m$. These results account for the
observed power-law distribution of the maximum amplitudes in acoustic
emission experiments of crystal plasticity, and are also broadly
applicable to other systems in the mean-field interface depinning
universality class, ranging from magnets to earthquakes.
\end{abstract}
\maketitle

Avalanche phenomena have been observed in a wide variety of disordered
systems that exhibit crackling noise near a depinning transition.
Examples include Barkhausen noise in soft magnetic
materials~\cite{Zapperi98,Sethna2001}, elastic depinning of charge
density waves~\cite{Lee79,Brazovskii04}, dynamics of
superconductors~\cite{Field95}, seismic activity in
earthquakes~\cite{Fisher98}, acoustic emission in mesoscopic crystal
plasticity~\cite{miguel2001idf}, and fracture
propagation~\cite{Zapperi99}. Although these varying materials/systems
have different microscopic details, on long length scales, the
statistical scaling behavior of avalanches appears to be universal. For
example, the distributions of avalanche sizes in sheared crystals and
in slowly magnetized soft magnets are both captured by the mean field
theory of a slowly driven elastic interface in a disordered
medium~\cite{Fisher98,Zaiser02,Brazovskii04,Doussal09,Uhl09,Tsekenis11}.

Recent experimental studies of slip avalanches in mesocopic crystal
plasticity have reported that the distribution of the maximum amplitude
$A_m$ of the acoustic emission (AE) signal from each avalanche follows
a power law $P(A_m) \sim A_m^{-\mu},$ where the exponent $\mu \approx
2$~\cite{Weiss97,Weiss2000,miguel2001idf,Weiss05,Weiss07,Fressengeas09}.
Since each avalanche contributes with only one maximum amplitude to the
histogram, many events are required to obtain good statistics for
$P(A_m)$. Thus the variations in the experimental values of $\mu$
depend on the experimental statistics. Owing to the proportionality
between the AE amplitude $A_m$ and the collective velocity $v_m$ of
dislocations~\cite{Weiss2000}, the distributions $P(A_m)$ and $P(v_m)$
should be characterized by the same scaling exponents and scaling
functions. So far, a theoretical prediction for the value of the
exponent $\mu$ has been lacking.

In this Letter, we present the first theoretical calculation of the
maximum velocity distribution, establishing a connection to the known
classes of extreme value statistics of correlated variables. In
particular, we derive the distribution of maximum velocities $P(v_m)$
from a mean field interface depinning model. We first show that the
probability distribution function (PDF) of the maximum velocity for
avalanches of fixed duration $T$ follows a universal scaling form
$P(v_m|T) = (2v_m T)^{-1/2}F(\sqrt{2v_m/T})$, with a scaling function
$F(x)$ that can be derived exactly by a mapping to an equivalent
problem of random excursions of Brownian motion in a logarithmic
potential. Although a general theory of extremal statistics for
strongly correlated variables is not known, much progress has been made
already for several classes of power-law correlated noise with an
$1/\omega^\alpha$ (where $\omega$ is the frequency) power spectrum.
Brownian noise corresponds to the particular case where
$\alpha=2$~\cite{Racz07,Burkhardt07}. The extreme value statistics of
power-law correlated noise typically have a robust scaling form, but
the scaling function depends on boundary conditions, the value from
which the maximum is measured, as well as other constraints on the time
evolution. For example, different scaling functions are obtained for
the maximum heights of periodic Gaussian interfaces: if the maximum is
measured relative to the spatially averaged height, the corresponding
EVS is determined by the so-called Airy distribution
function~\cite{Majumdar04,Majumdar05,Racz07,Rambeau11}, whereas
measuring the maximum relative to the boundary value leads to the
Rayleigh distribution~\cite{Rambeau09,Burkhardt07}. Here we demonstrate
that our problem of maximum heights of amplitudes of mean field
avalanches is equivalent to a related problem whose exact solution
obeys the same scaling form with a distinct function.  Finally, we show
that the overall distribution scales like $P(v_m)\sim v_m^{-2}$ by
integrating $P(v_m|T)$ against the duration PDF $F(T)$. The results of
this study are expected to be broadly applicable to plasticity,
earthquakes, Barkhausen noise in soft magnets and many other systems in
the mean field interface depinning universality class.

Our starting point is a zero-dimensional model of a slowly driven
elastic interface in a disordered medium, also known as the
Alessandro-Beatrice-Bertotti-Montorsi (ABBM) model~\cite{ABBM90},
corresponding to the dynamics of a particle pulled by an elastic spring
and an external field through a random force landscape. The position of
the particle $u(t)$ corresponds to the center-of-mass displacement of
the interface $u(t) = L^{-d}\int d^dx \hspace{1mm}u(\bm x,t)$, given
the local displacement $u(\bm x,t)$ at position $\bm x$ along the
interface length $L$ and time $t$ for an interface of dimension $d$
embedded in a $(d+1)$-dimensional space. In the ABBM model, the
evolution of the particle velocity $v= du/dt$ is obtained by a
time-differentiation of the mean field equation of motion of the
interface and given as~\cite{ABBM90,Colaiori08,Doussal09}
\begin{equation}\label{eq:ABBM}
\frac{dv}{dt} = -kv +c+\sqrt{v}\xi(t),
\end{equation}
where $c$ is the constant drift velocity, $k$ is the elastic coupling
constant and $\xi(t)$ is Gaussian white noise with autocorrelation
$\langle\xi(t)\xi(t')\rangle = 2D\delta(t-t')$ where $D$ is a constant
measure of disorder. In numerous studies, this model has been shown to
reproduce well the universal scaling laws for the size and duration
distributions near quasi-static depinning for systems with long-range
interactions~\cite{ABBM90,Zapperi98,Colaiori08,Doussal09,Papanikolaou11}.
It is relevant for the calculations below to recall the power law decay
of the distribution of avalanche durations $F_T(T)\approx
T^{-(2-c/D)}f_T\left(\frac{T}{T^*}\right)$, with a rate-dependent
exponent~\cite{ABBM90,Zapperi98,Colaiori08}, and such that, for $T\ll
T^*$ and $c\rightarrow0$, the distribution of durations follows the
mean field scaling law $F_T(T)\approx T^{-2}$. The exponential scaling
function $f_T(T/T^*)$ and the cutoff $T^*(k)$ to the scale invariance
can be computed analytically in the limit of
$c=0$~\cite{Colaiori08,Doussal11}. The probability distribution for the
velocity follows the Fokker-Planck equation
\begin{equation}\label{eq:FP}
\partial_tP(v,t)=\partial_v\left((kv-c)P(v,t)\right)+D\partial^2_v\left(vP(v,t)\right)
\end{equation}
which has a steady state solution given by
\begin{equation}\label{eq:PV_FK}
P(v) = v^{-1+\tilde c} \frac{\tilde k^{\tilde c}}{\Gamma(\tilde c)}e^{-\tilde k v},
\end{equation}
where $\Gamma(z)$ is the Gamma function, $\tilde c =c/D$ and $\tilde k
=k/D$~\cite{ABBM90,Zapperi98,Colaiori08,Doussal11}. The power-law
exponent depends linearly on the driving rate $c$, such that in the
adiabatic limit $c\rightarrow 0$, the distribution approaches the
well-known $v^{-1}$ scaling, which has been verified by experiments on
the dynamics of domain walls in ferromagnets~\cite{ABBM90_exp}.

\begin{figure}
\includegraphics[width=0.4\textwidth]{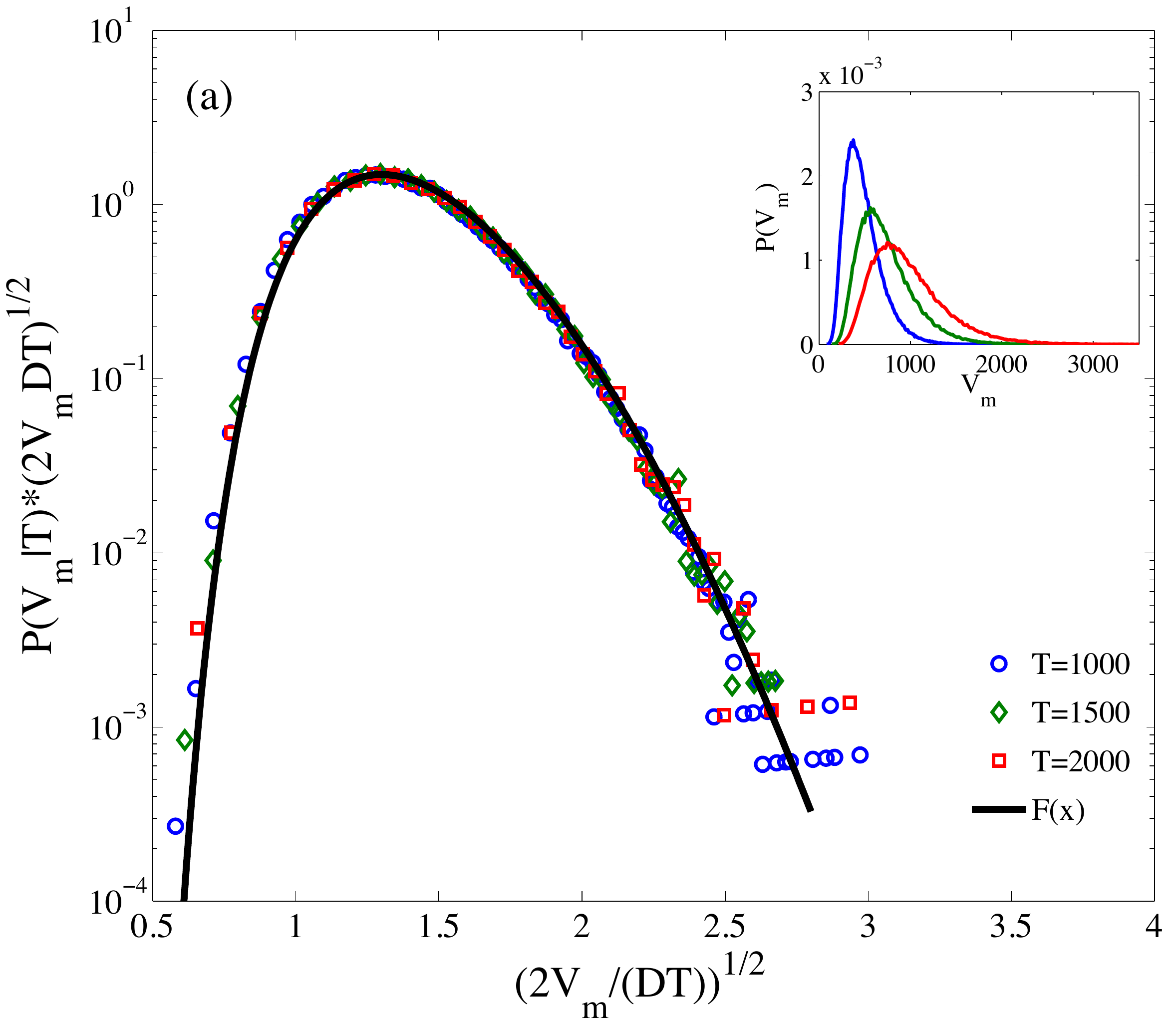}
\includegraphics[width=0.4\textwidth]{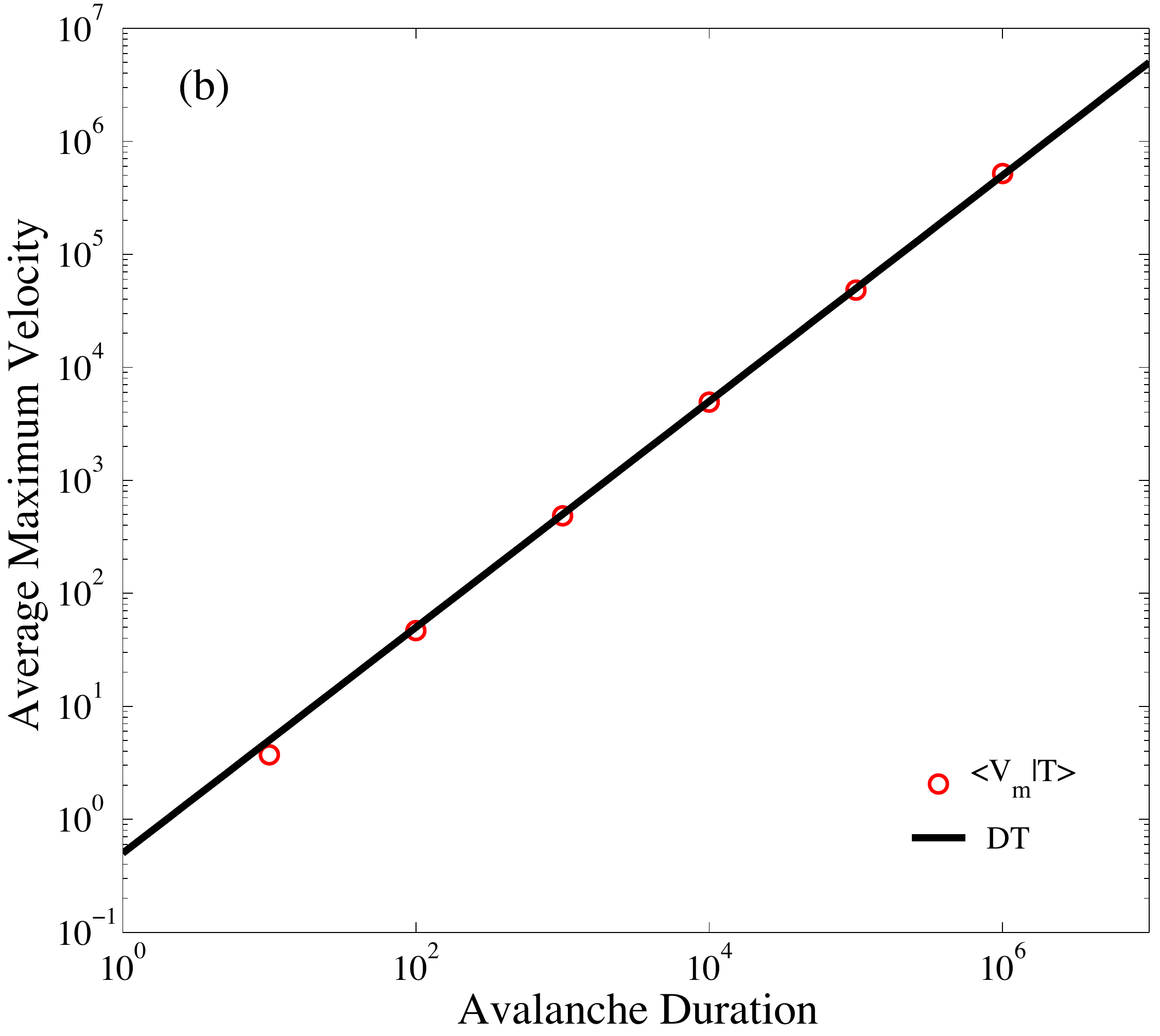}
\caption{ (Color online) (a) Data collapse of the PDF $P(v_m|T)$ from
numerical integration of Eq.~(\ref{eq:ABBM}) in the Ito interpretation
with parameter values $k=c=0$ and $D=1/2$. Large durations
($T\sim1000$) are required to obtain the scaling regime where
Eq.~(\ref{eq:Pvm_T}) holds. The collapse fits very well with the
analytically determined $F(x),$ which is represented by the solid line.
The inset figure shows the $P(v_m|T)$'s for different durations before
the rescaling. In panel (b), we show $\langle v_m|T\rangle $ as a
function of $T$, with the solid line representing the analytical
solution from Eq.~(\ref{eq:mean_Vm}) with $D=1/2.$} \label{fig:1}
\end{figure}

\medskip
\noindent {\it Maximum velocity distribution for avalanches of fixed
duration:-\/} By a change of variables to $x=2\sqrt{v}$,
Eq.~(\ref{eq:ABBM}) transforms to an additive-noise Langevin equation
$dx/dt= -kx/2+(2c-D)/x+\xi(t)$. The additional $1/x$ term comes from
the Ito interpretation of the multiplicative noise in
Eq.~(\ref{eq:ABBM}). This choice yields the correct Eq.~(\ref{eq:FP}).
Thus, in the adiabatic limit, near depinning, where $c\rightarrow 0$
and $k\rightarrow 0$, the velocity evolution can be mapped onto a
one-dimensional (1D) Brownian motion in a logarithmic potential. An
avalanche of duration $T$ corresponds to an excursion,  i.e. a path
$x(t)$ with $x(0)=x(T)=0$ and $x(t)>0$ for $0<t<T$. The extreme
displacement distribution for Brownian excursions can be derived using
the path integral formalism found in Refs.~\cite{Majumdar05,Rambeau09}.
We adapt this method to our problem, and determine the cumulative
distribution $C_{RW}(x_m|T)$ of the maximum displacement during
excursions for a Brownian motion in a logarithmic trap. The cumulative
distribution can be defined as $C_{RW}(x_m|T)=$
\begin{equation}\label{eq:CRW}
\lim_{\epsilon\rightarrow0}\frac{\int_{x(0)=\epsilon}^{x(T)=\epsilon}
\mathcal{D}xe^{-\int_0^Tdt\mathcal{L}_E}\prod_{t}\Theta(x(t))\Theta(x_m-x(t))
}{\int_{x(0)=\epsilon}^{x(T)=\epsilon}
\mathcal{D}xe^{-\int_0^Tdt\mathcal{L}_E}\prod_{t}\Theta(x(t))}
\end{equation}
where the Lagrangian is given by $\mathcal{L}_E=\frac{1}{4D}(\dot
x+\frac{1}{x})^2$.  The theta function products in the numerator
indicate that only paths that stay positive-valued between $t=0$ and
$t=T$ and have a maximum distance from the origin not greater than
$x_m$ are counted. The denominator is a normalization factor, counting
any excursion of duration $T$ without regard to its maximum value. The
Fokker-Planck equation (Eq.~(\ref{eq:FP})) with $c=k=0$ in terms of the
variable $x$ is Bessel's equation of order 1, thus the path integrals
from Eq.~(\ref{eq:CRW}) can be written as the matrix elements
$\langle\epsilon|\exp(-\hat HT)|\epsilon\rangle$ of the Hamiltonian
$\hat H=-\partial^2_x-\partial_x/x+1/x^2$ with appropriate boundary
conditions, and then expanded in terms of Bessel functions (details are
presented in~\cite{PRE}). From $C_{RW}(x_m|T)$, the PDF
$P(x_m|T)=\partial_{x_m}C(x_m|T)$ is determined. We find that the
$P(x_m|T)$ has the scaling form
\begin{equation}\label{eq:Px_T}
P(x_m|T) = \frac{1}{\sqrt{2DT}}F\left(\frac{x_m}{\sqrt{2DT}}\right),
\end{equation}
with scaling function
\begin{equation}\label{eq:Fx}
F(x) =
\frac{1}{x^5}\sum_{n=1}^\infty\frac{\lambda_n^2}{(J_2(\lambda_n))^2}\left[\frac{\lambda_n^2}{x^2}-4\right]e^{-\frac{\lambda_n^2}{2x^2}},
\end{equation}
where $\lambda_n$ is the $n$'th zero of the Bessel function $J_1(x).$
From Eq.~(\ref{eq:Px_T}), it also follows that the average maximum
displacement scales with duration as $T^{1/2}$, like the average
maximum relative heights of fluctuating interfaces~\cite{Majumdar04,
Majumdar05,Rambeau09}. Returning to the physical variable $v,$ we find
that the maximum velocity distribution in avalanches of fixed duration
$T$ has the scaling form
\begin{equation}\label{eq:Pvm_T}
P(v_m|T) = \frac{1}{\sqrt{2v_mDT}}F\left(\sqrt{\frac{2v_m}{DT}}\right).
\end{equation}
The average maximum velocity dependence on avalanche duration $T$ can
be obtained as the first moment of the conditional distribution
\begin{equation}\label{eq:mean_Vm}
\langle v_m|T\rangle = \int_0^\infty dv_m v_m P(v_m|T)=DT,
\end{equation}
where we used the fact that $\int dx x^2 F(x)=2.$

\begin{figure}
\includegraphics[width=0.4\textwidth]{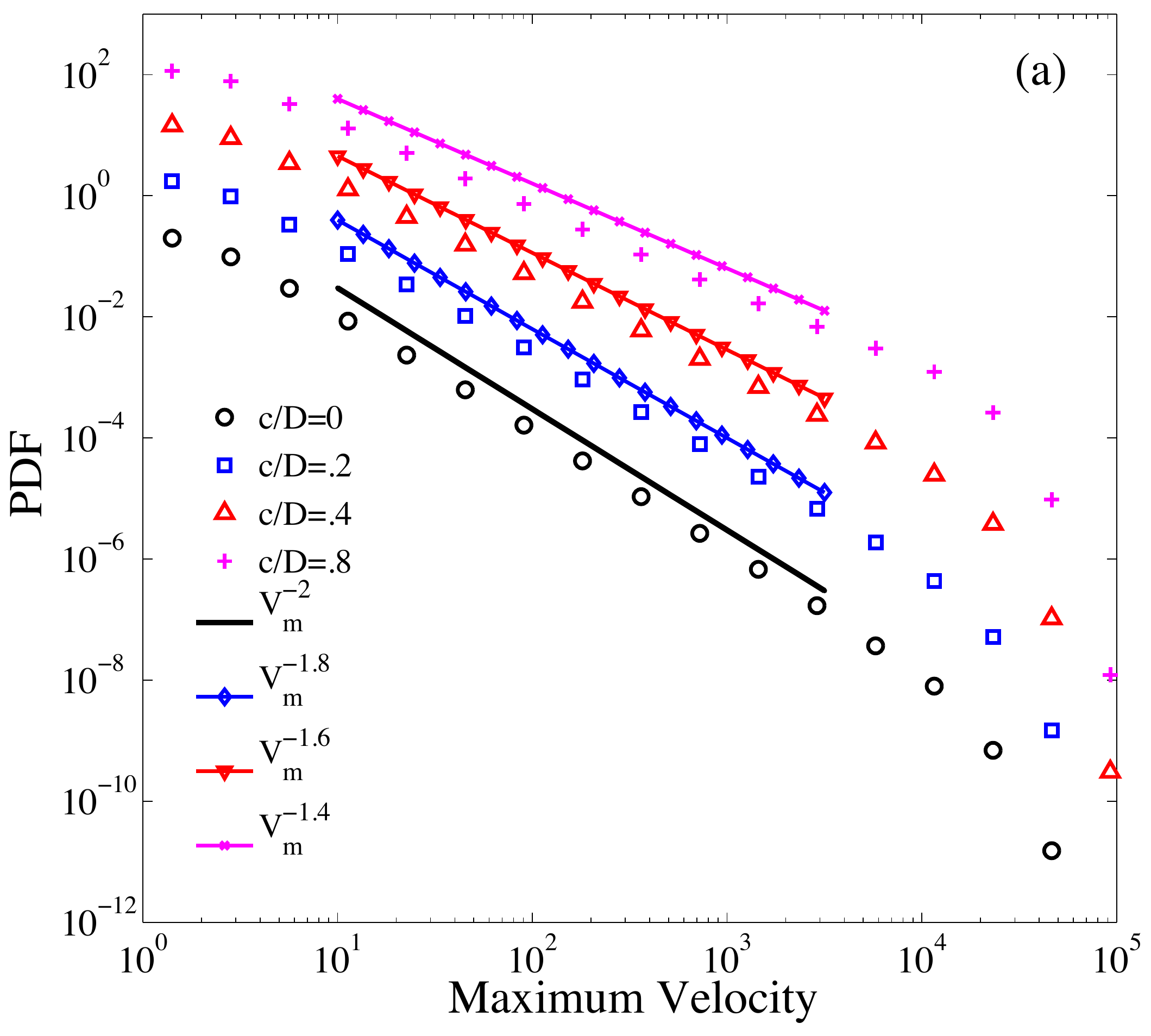}
\includegraphics[width=0.4\textwidth]{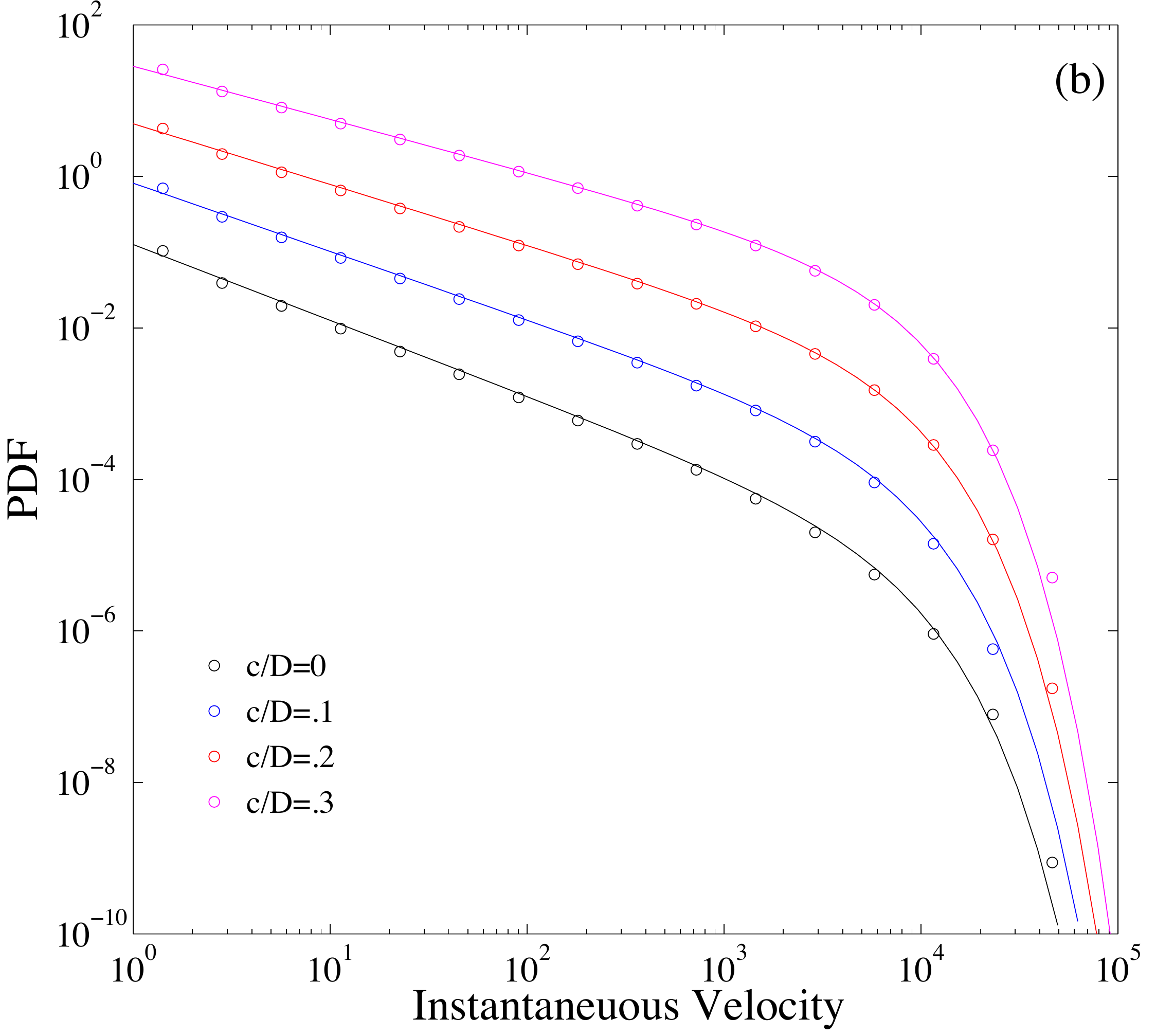}
\caption{(Color online) In panel (a), the PDF  $P(v_{m})$ obtained from
numerical integration of Eq.~(\ref{eq:ABBM}) is shown for several
values of $\tilde c$, all with $k=10^{-4}$ and $D=1/2.$ The PDFs are
offset vertically so they can be clearly distinguished. Above each
curve, a guideline is drawn indicating the power law analytically
predicted from Eq.~(\ref{eq:Pvm_2}). In panel (b), we show the instantaneous velocity PDF $P(v)$ for several values of $\tilde c.$ The solid lines represent the functional form predicted by Eq.~(\ref{eq:PV_FK}). Again, the PDFs are
offset for visibility.} \label{fig:2}
\end{figure}

Using the same method, we also determine that the PDF of the
instantaneous velocity $v$ at time $t$ in an avalanche of duration $T$
is given by
\begin{equation}\label{eq:Pv_T}
P(v,t|T) =v\left(\frac{T}{Dt(T-t)}\right)^2e^{-vT/(Dt(T-t))}.
\end{equation}
The first moment of $P(v,t|T)$ gives $\langle v(t)|T\rangle =
2Dt(T-t)/T,$ which is the parabolic average avalanche shape discussed in
\cite{Sethna2001,Mehta02,Kuntz00}.

We have verified Eqs.~(\ref{eq:Pvm_T}) and (\ref{eq:mean_Vm})
numerically  by integrating Eq~(\ref{eq:ABBM}) (see Fig~(\ref{fig:1})).
Large durations ($T\sim1000$) must be explored for the scaling function
to converge to the one predicted by our continuum derivation, but the
results are in accord with predictions. We obtained improved
statistical results for data collapse to the same scaling function
using the computationally-efficient discrete velocity shell model
\cite{Fisher98,Kuntz00}, which obtains its scaling regime at smaller
durations.  These results will be reported elsewhere \cite{PRE}.

Although our analytical calculation was performed exactly only for
$k=c=0,$ the scaling form in Eq.~(\ref{eq:Pvm_T}) gives a good collapse
of simulation data away from criticality as well, even when including
durations $T\sim T^*(k)$~\cite{PRE}. Therefore the dependence of the
scaling form on $k$ is likely to be weak. Since the driving rate
parameter $\tilde c$ is dimensionless, one might anticipate that
nonzero values of the driving rate $c$ modify the scaling function, but
not the scaling form. Indeed, the exact form of the modification can be
calculated analytically with a slight generalization of the above
calculation \cite{PRE}.

\medskip
\noindent {\it Maximum and instantaneous velocity statistics:-\/} We
now investigate the maximum velocity distribution integrated over all
durations. This distribution is equivalent to the $P(A)$ of the maximum
AE amplitude, $A$, deduced from the time series in AE experiments on
crystal plasticity. From Eq.~(\ref{eq:Pv_T}), we can determine the PDF
of the maximum avalanche velocity $P(v_m)$ by integrating $P(v_m|T)$
over avalanche durations $T$ weighted by their distribution $F_T(T)\sim
T^{-2+\tilde c}$, for $T\ll T^*$. Our numerics indicate that $P(v_m|T)$
satisfies Eq.~(\ref{eq:Pvm_T}) at least for durations $T<T^*$. Thus,
the distribution of $P(v_m)$ is
\begin{eqnarray}\label{eq:Pvm}
P(v_m) &\sim& \int_0^{T^*}\frac{dT}{T^{2-\tilde c}}P(v_m|T),
\end{eqnarray}
and near depinning, where $T^*\rightarrow \infty$,  we have
\begin{eqnarray}\label{eq:Pvm_2}
P(v_m) &\sim& v_m^{-2+\tilde c}.
\end{eqnarray}
Similarly, we obtain the PDF $P(v)$ for the instantaneous velocity $v$
at an arbitrary time by integrating $P(v,t|T)$ over the time spent in
an avalanche of duration $T$ and then over the distribution of
durations, giving
\begin{eqnarray}\label{eq:Pv}
P(v)&=&\int_0^{T^*} dT \int_0^T dt P(v,t|T)F(T)\\&\sim&v\int_0^{T*} \frac{dT}{T^{3-\tilde c}}G(v/DT)
\end{eqnarray}
where $G(x)=\int_0^1du(u(1-u))^{-2}\exp(-x(u(1-u))^{-1}).$ $G(x)\sim
x^{-1}$ for $x\ll1$ and decays exponentially for $x\gg 1,$ so in the
limit $T^*\rightarrow\infty$, we recover the $P(v)\sim v^{-1+\tilde c}$
scaling predicted by the steady state equation for $\tilde c
<1$~\cite{ABBM90,Colaiori08,Doussal09,Doussal11}.  In
Fig.~(\ref{fig:2}), we show numerically calculated PDFs $P(v_m)$ and
$P(v)$ for various values of $c.$ The distributions agree with the
predictions of Eqs.~(\ref{eq:Pvm_2}) and~(\ref{eq:PV_FK}).

In addition to the exponents, it would be interesting to measure
the predicted scaling form of the $P(v_m|T)$ over fixed durations from
AE experiments, in the corresponding regime where the distribution of
maximum amplitudes $P(A_m)\sim A_m^{-2}$ was
observed~\cite{Weiss97,Weiss2000,miguel2001idf,Weiss05,Weiss07,Fressengeas09}.
The distribution $P(v_m|T)$ was calculated exactly only at the
depinning transition with $k=c=0$, but numerical evidence strongly
suggests that an indistinguishable scaling form occurs away from the
transition, whose dependence on the elastic
coupling constant and the driving rate still needs to be studied in
more detail.  Finally, what happens beyond mean field theory
remains an open question, despite the apparently good agreement of our
calculations with available experimental data. For instance, it is
unclear to us whether the exponent $\mu$ in $P(v_m)=v_m^{-\mu}$ is
expressible in terms of the other avalanche exponents ($\tau$,
$\sigma\nu z$, etc.) or if it is independent.

\noindent {\it Acknowledgments:} L.\,A. is grateful for support from
the Center of Excellence for Physics of Geological Processes. This work
was partially supported by the National Science Foundation through
grants DMR-1005209, and DMS-1069224.

\bibliographystyle{apsrev4-1}
\bibliography{references}

\end{document}